# Is the Missing Ultra-Red Material Colorless Ice?

W.M. Grundy[1]

1. Lowell Observatory, 1400 W. Mars Hill Rd., Flagstaff AZ 86001.

Primary contact:     Will Grundy
Mailing address:     Lowell Observatory, 1400 W. Mars Hill Rd., Flagstaff AZ 86001
E-mail:              W.Grundy@lowell.edu
Voice:               928-233-3231
Fax:                 928-774-6296

**ABSTRACT**

The extremely red colors of some transneptunian objects and Centaurs are not seen among the Jupiter family comets which supposedly derive from them. Could this mismatch result from sublimation loss of colorless ice? Radiative transfer models show that mixtures of volatile ice and non-volatile organics could be extremely red, but become progressively darker and less red as the ice sublimates away.

**1. Introduction**

Redder visual wavelength colors have been observed for asteroids at larger heliocentric distances (e.g., Tholen 1984; Zellner et al. 1985; Hartmann et al. 1987). That distribution of coloration has been attributed to an increasing proportion of reddish organic materials relative to silicates incorporated into bodies formed in colder protoplantary disk environments further from the Sun (e.g., Gradie and Veverka 1980; Vilas and Smith 1985). More recently, the trend has been shown to extend to the outer Solar System, with numerous primitive bodies in the giant-planet-crossing Centaur region and beyond revealed to have extremely red colors. They are not all red, however. Some Centaurs show more neutral colors (Peixinho et al. 2003; Tegler et al. 2008) and the distribution of colors among transneptunian objects (TNOs) is very broad (Peixinho et al. 2004; Doressoundiram et al. 2005, 2008).

From dynamical simulations, TNOs are thought to be the source of Centaurs and thence Jupiter-family comets (JFCs; Duncan and Levison 1997; Lowry et al. 2008). Jewitt (2002) noted that despite this widely-accepted genetic link, the visual wavelength colors of TNOs, Centaurs, and JFCs are mutually inconsistent and concluded that an ultra-red material must be destroyed, removed, or buried as objects approach the Sun. Complex organics such as tholins (e.g., Cruikshank et al. 2005; de Bergh et al. 2008) can be extremely red, and are considered to be likely ingredients of outer Solar System bodies, since they are readily produced by energetic radiation acting on cosmochemically abundant volatiles. However, the tholins described in the literature are not volatile at room temperature and so would not sublimate away as an object migrates from the transneptunian region to the inner Solar System. Jewitt (2002) showed that cometary activity could bury a radiolytic surface veneer. But if the ultra-red material is representative of bulk composition, such resurfacing might not actually change the surface color. The discrete clustering of colors reported by Barucci et al. (2005), the absence of color dependence on object size (e.g., Doressoundiram et al. 2008), the rarity of confirmed color lightcurve variations (e.g., Jewitt and Luu 2001; Doressoundiram et al. 2005), and the highly correlated colors of small, captured binaries (Benecchi et al. 2008) are all consistent with colors being related to primordial bulk composition, rather than being just an environmental surface effect. Small, red, organic particles were probably abundant during the accretion of TNOs (e.g., Greenberg 1998; Hanner and Bradley 2004; Nakamura-Messenger et al. 2006; Sandford et al. 2006; Alexander et al. 2007) and probably contributed to their bulk compositions.

This paper offers an explanation for the loss of red coloration which does not require the ultra-red material to be confined to a thin surface layer: sublimation loss of ice from a mixture with red material. Grundy and Stansberry (2003) noted that under certain circumstances, adding a transparent material such as ice to a tholin could make its visual reflectance spectrum much more red. This paper builds on that idea to explore the seemingly paradoxical possibility that the ultra-red coloring agent remains present throughout the transition from the Kuiper belt to the inner Solar System, with the observed color change resulting from the loss of volatile ice. Ice is effectively colorless at visual wavelengths, but under certain circumstances, a transparent material can dramatically enhance the effect of another coloring agent.

## 2. Models

How could adding a colorless, transparent material to a strongly colored material increase the reflected color contrast? It all depends on texture. A solid absorbs light passing through it according to the Beer-Lambert law,

$$I(\lambda) = I_0(\lambda) e^{-l\alpha(\lambda)}, \tag{1}$$

where $I_0(\lambda)$ is the incident flux at wavelength $\lambda$, $I(\lambda)$ is the transmitted flux, $l$ is the distance traversed within the material, and $\alpha(\lambda)$ is the Lambert absorption coefficient of the material, with units of inverse length. Texture is a major influence on the mean optical path length $l_{MOPL}$ (e.g., Clark and Lucey 1984), the average distance photons travel within a material prior to being scattered out of it. Imagine a surface composed of a material with different absorption coefficients at two wavelengths. If the texture of the surface is such that the product $l_{MOPL}\alpha(\lambda)$ is much greater that unity at both wavelengths, little light will escape at either wavelength. The surface will look black, not colorful. Likewise, if $l_{MOPL}\alpha(\lambda) \ll 1$ at both wavelengths, little light will be absorbed at either wavelength and the surface will look white, not colorful. To maximize the colorfulness, the contrast in reflectance between the two wavelengths, $l_{MOPL}\alpha(\lambda)$ at the two wavelengths should straddle an intermediate range between these two extremes.

The addition of a colorless material can shift a surface from the $l_{MOPL}\alpha(\lambda) \gg 1$ (black) regime into the $l_{MOPL}\alpha(\lambda) \approx 1$ (colorful) regime if the absorbing material is dispersed within the colorless material like a pigment mixed into a white paint base. This can be done by increasing the spatial density of scatterers to reduce $l_{MOPL}$, or by diluting the pigment to reduce $\alpha(\lambda)$. This paper focuses on the latter mechanism. Dilution of $\alpha(\lambda)$ occurs when pigment particles are dispersed at the spatial scale of the wavelength or smaller. Grundy and Stansberry (2003) simulated this effect for tholin diluted in water ice using a trivial mixing model, a volume-weighted average of optical constants.

In this paper, I use Maxwell-Garnett effective medium theory (e.g., Garnett 1904; Niklasson et al. 1981) to simulate mixtures of ice with carbonaceous particles much smaller than the wavelength. Maxwell-Garnett theory has previously been used to model the reddening effect of nanophase iron particles in space-weathered silicates (Hapke 2001). Sub-micron carbonaceous and organic particles are familiar from primitive meteorites, cometary dust, and interplanetary dust (e.g., Nakamura-Messenger et al. 2006; Sandford et al. 2006; Alexander et al. 2007) so it is not unreasonable to expect particles such as these to have been incorporated into TNOs.

Maxwell-Garnett effective medium theory approximates the effect of sub-wavelength inclusions by means of effective bulk optical constants for the inclusions plus surrounding matrix, given by

$$\varepsilon_{\text{eff}}(\lambda) = \varepsilon_{\text{m}}(\lambda) \frac{\varepsilon_{\text{i}}(\lambda) + 2\varepsilon_{\text{m}}(\lambda) + 2f_{\text{i}}(\varepsilon_{\text{i}}(\lambda) - \varepsilon_{\text{m}}(\lambda))}{\varepsilon_{\text{i}}(\lambda) + 2\varepsilon_{\text{m}}(\lambda) + f_{\text{i}}(\varepsilon_{\text{m}}(\lambda) - \varepsilon_{\text{i}}(\lambda))} , \qquad (2)$$

where $\varepsilon(\lambda)$ is the relative permittivity (also known as the dielectric constant), subscript 'i' refers to the included material, subscript 'm' refers to the matrix, and $f_i$ is the volume fraction occupied by inclusions. Various readily-convertible forms of optical constants are encountered in the literature. Relative permittivity $\varepsilon(\lambda)$ is the square of the complex refractive index $n(\lambda) + ik(\lambda)$. The Lambert absorption coefficient $\alpha(\lambda)$ equals $4\pi k(\lambda)/\lambda$. The models in this paper made use of optical constants for water ice from Warren (1984, for the real part) and from Grundy and Schmitt (1998, for the imaginary part). Optical constants for a representative tholin were taken from Khare et al. (1984).

Two types of larger, aggregate particles were simulated to explore the diversity of possible configurations as ice is lost from a mixture with small carbonaceous particles. In model 1, the aggregates consist of a constant volume of carbonaceous material (equivalent to the volume of a 5 μm diameter sphere) plus a variable amount of H₂O ice, leading to aggregate sizes dependent on H₂O ice content. The majority species was treated as the matrix, with the minority species as subwavelength inclusions. In model 2, the ratio of H₂O ice to carbonaceous material was allowed to vary, with both materials being treated as subwavelength inclusions in a vacuum matrix. In this model, the aggregate particle size was held constant at 20 μm to represent scattering from larger voids or cracks within the bulk material. Such larger scale defects could result from a process such as impact gardening. In both models, the carbonaceous material was taken to consist of 99% tholin particles and 1% carbon particles (optical constants from Jäger et al. 1998). The carbon was added to reduce the albedo of the pure carbonaceous end-member. Hapke scattering theory (Hapke 1993) was used to account for multiple scattering by the aggregate particles. Fig. 1 shows example disk-integrated albedo spectra from the two models.

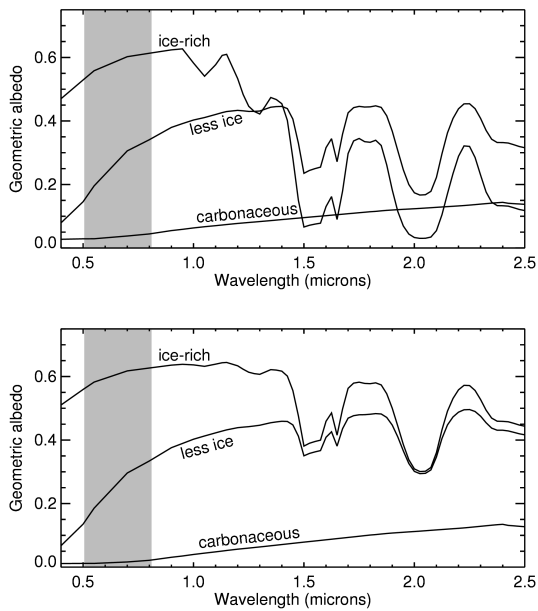

**Fig. 1:** Simulated disk-integrated albedo spectra of bodies covered with model 1 aggregates (top panel) and model 2 aggregates (bottom panel). The shaded area indicates the wavelength range of standard V and R filters. In both models, the color at these wavelengths is reddest for intermediate compositions. The most obvious difference between models 1 and 2 is that the infrared H₂O ice absorption bands are much deeper in ice-rich cases with model 1, because the particle size is larger and the mean optical path lengths within H₂O are longer.

For both models, the color at V and R band wavelengths is redder for intermediate compositions than it is for either ice-rich or carbonaceous compositions. As ice is lost from these mixtures, their visual appearance progresses from a nearly-white, ice-rich composition, through very red colors at intermediate compositions, and finally to black. The color trend is illustrated in more detail in the top panel of Fig. 2, which shows the V−R color in units of magnitudes relative to Solar as a function of ice fraction. The maximum redness and the mixing ratio which produces it differ considerably between the two models, but the general behavior of maximum red-ness at intermediate rather than pure-carbonaceous compositions is consistent, despite the different configurations. For an ice-dominated initial composition, removing ice results in an initial increase in redness, but as less and less ice remains, the redness inevitably declines. Loss of ice also results in rapidly diminishing albedos and weakening near-infrared water ice absorption bands, as shown in the lower two panels of Fig. 2. Strong $H_2O$ ice absorption bands are only really evident when the composition is nearly pure ice.

## 3. Discussion

As an explanation for the loss of red coloration as TNOs evolve closer to the Sun, loss of a ubiquitous volatile material such as ice seems more plausible than loss of an as-yet undiscovered red volatile material or of non-volatile, complex, red organics.

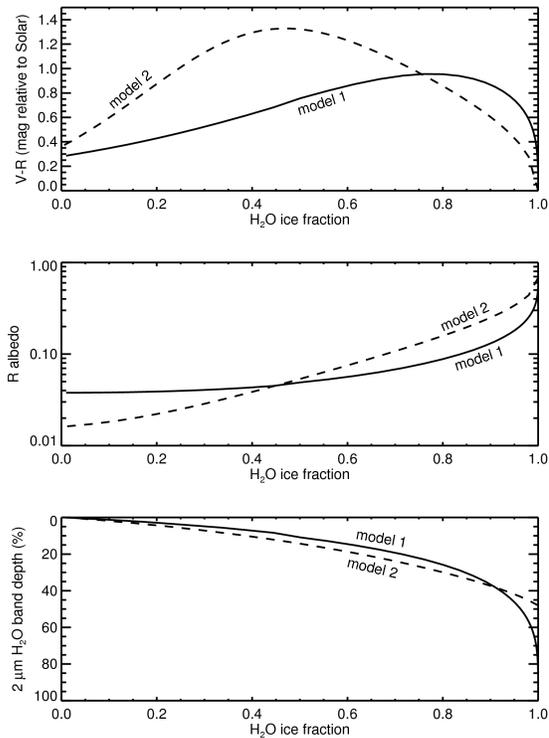

**Fig. 2:** The top panel shows the *V–R* color in units of magnitudes relative to Solar as a function of ice fraction. Higher values are more red. The middle panel shows the *R*-band geometric albedo. The bottom panel shows the fractional depth of the strong $H_2O$ ice absorption band at 2 μm, calculated as one minus the ratio of the band center albedo (measured at 2.03 μm) divided by the mean continuum albedo (measured as the average of albedos at 1.83 and 2.23 μm).

In addition to the loss of redness from TNOs through Centaurs to JFCs, object classes closer to the Sun have lower average albedos (e.g., Fernández et al. 2003; Lamy et al. 2004; Brucker et al. 2008; Stansberry et al. 2008). The models proposed in this paper duplicate this behavior nicely, with visual albedos declining as ice is lost and reddening diminishes. Models involving burial of red, radiolytic veneers by cometary activity (e.g., Jewitt 2002; Delsanti et al. 2004) may have difficulty explaining this drop in albedo, if irradiated exteriors are expected to be darker than less-irradiated interiors (e.g., Thompson et al. 1987; Moroz et al. 2004).

The two models presented here by no means encompass the universe of plausible configurations or compositions. Their purpose is simply to explore the general color and albedo behavior as red organic materials are combined at sub-wavelength scales with transparent ices, and to illustrate the sensitivity to model assumptions. The Khare et al. (1984) Titan analog tholin represents a diverse family of complex organics (e.g., Cruikshank et al. 2005; de Bergh et al. 2008). Other materials with red visual spectra would behave similarly in our models. The same can be said for the Jäger et al. (1998) carbon sample. That particular specimen was produced by pyrolysis of cellulose at 800 °C, but other forms of carbon including soot, graphite, and charcoal share the property of high opacity throughout the visible and near-infrared spectral region, which is all that really matters for present purposes. Water ice is also not the only cosmochemically abundant ice which is transparent at visual wavelengths. Other, more volatile ices such as $CO_2$, $H_2CO$, $CH_3OH$, and $CH_4$ are plausible constituents of TNOs (e.g., de Bergh et al. 2008) and could play similar roles to that of H$_2$O in the models presented in the previous section. Being more volatile than water, they would sublimate away at even greater heliocentric distances. If multiple ices were involved, each would sublimate away at a characteristic range of distances, and an inventory of the surviving ices at the surface of an object could place constraints on how close it had approached the Sun.

The most consistently red outer Solar System population of all is the dynamically "cold" core of the classical Kuiper belt (e.g., Tegler et al. 2003; Doressoundiram et al. 2005, 2008; Gulbis et al. 2006). Cold classical Kuiper belt objects (CCKBOs) occupy near-circular, low-inclination, non-resonant orbits. In addition to their red colors, they differ from other TNOs in their magnitude-frequency distribution (Levison and Stern 2001; Petit et al. 2008), higher rate of binarity (Noll et al. 2008), and higher average albedo (Grundy et al. 2005; Brucker et al. 2008). Of all TNOs, CCKBOs are thought to have formed furthest from the Sun (e.g., Levison et al. 2008). They could represent the far edge of a compositionally-zoned protoplanetary nebula, nearer regions of which are sampled by various other relict populations. Although CCKBOs may not be the direct source of Centaurs and JFCs, with their red colors and higher albedos, CCKBO surfaces could consist of ice plus carbonaceous mixtures as discussed in this paper, but with higher ice abundances than seen in other, more thermally evolved populations. The models presented here offer a reasonable explanation for the CCKBOs' combination of high albedos, red colors, and non-detection of ice absorption bands in low signal precision spectra. If their surfaces are indeed composed of ice plus carbonaceous mixtures as discussed in this paper, higher signal precision near-infrared spectroscopy should be able to detect the ices via subtle near-infrared absorption bands.

Redder Centaurs have higher average albedos than those with grayer coloration (Tegler et al. 2008), consistent with the scenario described here if red Centaurs are fresh arrivals from the Kuiper belt while gray ones have experienced warmer temperatures closer to the Sun and lost most of their surface ice. Whether or not this picture is consistent with the absence of statistically significant correlations between Centaur orbital elements and colors (Tegler et al. 2008) remains to be explored. One might also expect this mechanism to produce a continuum of Centaur colors rather than their observed bimodal pattern (e.g., Peixinho et al. 2003; Tegler et al. 2003). However, a bimodal color distribution would be produced if most ice is lost over a narrow range of heliocentric distances or if the distribution of minimum heliocentric distances sampled by Centaurs was itself bimodal.

The existence of gray TNOs at heliocentric distances where ice is not volatile may require other mechanisms to produce their coloration. Possibilities include early thermal activity and compositional zoning in the protoplanetary disk, the results of which may ultimately be linkable to distinct compositional classes of comets (e.g., A'Hearn et al. 1995). That ultra-red colors are not seen among the regular satellites of the giant planets is consistent with their surfaces having very different micro-structures from red TNOs, perhaps because the TNOs formed in thermal and chemical environments very different from the circumplanetary nebulae where the satellites formed (e.g., Simonelli et al. 1989).

As small bodies evolve from the Kuiper belt toward the inner Solar System, their diminishing red colors and declining albedos could result from sublimation loss of ices. If ices and complex red organics are indeed mixed at the sub-wavelength scale on the surfaces of TNOs as described in this paper, their surfaces will darken and become less red if the surface ice sublimates away, as it inevitably will for a body approaching the Sun.  However, reality is always more complex.  Additional processes are also surely at work on the surfaces of TNOs evolving into comets, perhaps most significantly those associated with cometary activity itself (e.g., Jewitt 2002; Delsanti et al. 2004).  Additionally, energetic impacts disrupt surface and internal structures and fragment larger bodies into smaller ones (typical comet nuclei are much smaller than known TNOs).  Radiolysis and photolysis break chemical bonds and enable production of new compounds, potentially including complex organics (e.g., Johnson et al. 1987; Cooper et al. 2003; Moroz et al. 2004; Hudson et al. 2008).  Sputtering removes ice and other molecules from the surface (e.g., Johnson 1998).  A detailed analysis of the rates and efficiencies of these processes could help elucidate which of them dominate under specific circumstances.

**Acknowledgments**

This work was partially supported by NASA Planetary Geology and Geophysics grant NNG04G172G to Lowell Observatory.  Thanks are also due to two anonymous reviewers and to the free and open source software communities for providing key tools used to complete this project, notably Linux, the GNU tools, OpenOffice.org, and FVWM.

**REFERENCES**

A'Hearn, M.F., R.L. Millis, D.G. Schleicher, D.J. Osip, and P.V. Birch 1995.  The ensemble properties of comets: Results from narrowband photometry of 85 comets, 1976-1992.  *Icarus* **118,** 223-270.

Alexander, C.M.O'D., A.P. Boss, L.P. Keller, J.A. Nuth, and A. Weinberger 2007. Astronomical and meteoritic evidence for the nature of interstellar dust and its processing in protoplanetary disks. In: B. Reipurth, D. Jewitt, K. Keil (Eds.), *Protostars and Planets V*, Univ. of Arizona Press, Tucson, 801-813.

Barucci, M.A., I.N. Belskaya, M. Fulchignoni, and M. Birlan 2005.  Taxonomy of Centaurs and trans-neptunian objects. *Astron. J.* **130,** 1291-1298.

Benecchi, S.D., K.S. Noll, W.M. Grundy, M.W. Buie, D. Stephens, and H.F. Levison 2008.  The correlated colors of transneptunian binaries.  *Icarus* (in press, see http://arxiv.org/abs/0811.2104).

Brucker, M.J., W.M. Grundy, J.A. Stansberry, J.R. Spencer, S.S. Sheppard, E.I. Chiang, and M.W. Buie 2008.  High albedos of low inclination classical Kuiper belt objects. *Icarus* (submitted).

Clark, R.N., and P.G. Lucey 1984.  Spectral properties of ice-particulate mixtures and implications for remote sensing: 1. Intimate mixtures.  *J. Geophys. Res.* **89,** 6341-6348.

Cooper, J.F., E.R. Christian, J.D. Richardson, and C. Wang 2003.  Proton irradiation of centaur, Kuiper belt, and Oort cloud objects at plasma to cosmic ray energy.  *Earth, Moon, and Planets* **92,** 261-277.

Cruikshank, D.P., H. Imanaka, and C.M. Dalle Ore 2005.  Tholins as coloring agents on outer Solar System bodies.  *Adv. Space Res.* **36,** 178-183.

de Bergh, C., B. Schmitt, L.V. Moroz, E. Quirico, and D.P. Cruikshank 2008. Laboratory data on ices, refractory carbonaceous materials, and minerals relevant to transneptunian objects and Centaurs. In: A. Barucci, H. Boehnhardt, D. Cruikshank, A. Morbidelli (Eds.), *The Solar System Beyond Neptune*, Univ. of Arizona Press, Tucson, 483-506.

Delsanti, A., O. Hainaut, E. Jourdeuil, K.J. Meech, H. Boehnhardt, and L. Barrera 2004. Simultaneous visible-near IR photometric study of Kuiper Belt Object surfaces with ESO/Very Large Telescopes. *Astron. & Astrophys.* **417,** 1145-1158.

Doressoundiram, A., N. Peixinho, C. Doucet, O. Mousis, M.A. Barucci, J.M. Petit, and C. Veillet 2005. The Meudon multicolor survey (2MS) of centaurs and trans-neptunian objects: Extended dataset and status on the correlations reported. *Icarus* **174,** 90-104.

Doressoundiram, A., H. Boehnhardt, S.C. Tegler, and C. Trujillo 2008. Color properties and trends of the transneptunian objects. In: A. Barucci, H. Boehnhardt, D. Cruikshank, A. Morbidelli (Eds.), *The Solar System Beyond Neptune*, Univ. of Arizona Press, Tucson, 91-104.

Duncan, M.J., and H.F. Levison 1997. A scattered disk of icy objects and the origin of Jupiter-family comets. *Science* **276,** 1670-1672.

Fernández, Y.A., S.S. Sheppard, and D.C. Jewitt 2003. The albedo distribution of jovian Trojan asteroids. *Astron. J.* **126,** 1563-1574.

Garnett, J.C.M. 1904. Colours in metal glasses and in metallic films. *Phil. Trans. Roy. Soc.* **203,** 385-420.

Gradie, J., and J. Veverka 1980. The composition of the Trojan asteroids. *Nature* **283,** 840-842.

Greenberg, J.M. 1998. Making a comet nucleus. *Astron. & Astrophys.* **330,** 375-380.

Grundy, W.M., and B. Schmitt 1998. The temperature-dependent near-infrared absorption spectrum of hexagonal $H_2O$ ice. *J. Geophys. Res.* **103,** 25809-25822.

Grundy, W.M., and J.A. Stansberry 2003. Mixing models, colors, and thermal emissions. *Earth, Moon, and Planets* **92,** 331-336.

Grundy, W.M., K.S. Noll, and D.C. Stephens 2005. Diverse albedos of small trans-neptunian objects. *Icarus* **176,** 184-191.

Gulbis, A.A.S., J.L. Elliot, and J.F. Kane 2006. The color of the Kuiper belt core. *Icarus* **183,** 168-178.

Hanner, M.S., and J.P. Bradley 2004. Composition and mineralogy of cometary dust. In: M.C. Festou, H.U. Keller, H.A. Weaver (Eds.), *Comets II*, Univ. of Arizona Press, Tucson, 555-564.

Hapke, B. 1993. *Theory of reflectance and emittance spectroscopy*. Cambridge Univ. Press, New York.

Hapke, B. 2001. Space weathering from Mercury to the asteroid belt. *J. Geophys. Res.* **106,** 10039-10073.

Hartmann, W.K., D.J. Tholen, and D.P. Cruikshank 1987. The relationship of active comets,

"extinct" comets, and dark asteroids. *Icarus* **69,** 33-50.

Hudson, R.L., M.E. Palumbo, G. Strazzulla, M.H. Moore, J.F. Cooper, and S.J. Sturner 2008. Laboratory studies of the chemistry of transneptunian object surface materials. In: A. Barucci, H. Boehnhardt, D. Cruikshank, A. Morbidelli (Eds.), *The Solar System Beyond Neptune*, Univ. of Arizona Press, Tucson, 507-523.

Jäger, C., H. Mutschke, and T. Henning 1998. Optical properties of carbonaceous dust analogues. *Astron. & Astrophys.* **332,** 291-299.

Jewitt, D.C., and J.X. Luu 2001. Colors and spectra of Kuiper belt objects. *Astron. J.* **122,** 2099-2114.

Jewitt, D.C. 2002. From Kuiper belt object to cometary nucleus: The missing ultrared matter. *Astron. J.* **123,** 1039-1049.

Johnson, R.E. 1998. Sputtering and desorption from icy surfaces. In: B. Schmitt, C. de Bergh, M. Festou (Eds.), *Solar System Ices*, Kluwer Academic Publishers, Boston, 303-334.

Johnson, R.E., J.F. Cooper, L.J. Lanzerotti, and G. Strazzulla 1987. Radiation formation of a non-volatile comet crust. *Astron. & Astrophys.* **187,** 889-892.

Khare, B.N., C. Sagan, E.T. Arakawa, F. Suits, T.A. Callcott, and M.W. Williams 1984. Optical constants of organic tholins produced in a simulated Titanian atmosphere: From soft X-ray to microwave frequencies. *Icarus* **60,** 127-137.

Lamy, P.L., I. Toth, Y.R. Fernández, and H.A. Weaver 2004. The sizes, shapes, albedos, and colors of cometary nuclei. In: M.C. Festou, H.U. Keller, H.A. Weaver (Eds.), *Comets II*, Univ. of Arizona Press, Tucson 223-264.

Levison, H.F., and S.A. Stern 2001. On the size dependence of the inclination distribution of the main Kuiper Belt. *Astron. J.* **121,** 1730-1735.

Levison, H.F., A. Morbidelli, C. VanLaerhoven, R. Gomes, and K. Tsiganis 2008. Origin of the structure of the Kuiper belt during a dynamical instability in the orbits of Uranus and Neptune. *Icarus* **196,** 258-273.

Lowry, S., A. Fitzsimmons, P. Lamy, and P. Weissman 2008. Kuiper belt objects in the planetary region: The Jupiter-family comets. In: A. Barucci, H. Boehnhardt, D. Cruikshank, A. Morbidelli (Eds.), *The Solar System Beyond Neptune*, Univ. of Arizona Press, Tucson, 397-410.

Moroz, L., G. Baratta, G. Strazzulla, L. Starukhina, E. Dotto, M.A. Barucci, G. Arnold, and E. Distefano 2004. Optical alteration of complex organics induced by ion irradiation: 1. Laboratory experiments suggest unusual space weathering trend. *Icarus* **170,** 214-228.

Nakamura-Messenger, K., S. Messenger, L.P. Keller, S.J. Clemett, and M.E. Zolensky 2006. Organic globules in the Tagish Lake meteorite: Remnants of the protosolar disk. *Science* **314,** 1439-1442.

Niklasson, G.A., C.G. Granqvist, and O. Hunderi 1981. Effective medium models for the optical properties of inhomogeneous materials. *Appl. Opt.* **20,** 26-30.

Noll, K.S., W.M. Grundy, D.C. Stephens, H.F. Levison, and S.D. Kern 2008. Evidence for two

populations of classical transneptunian objects: The strong inclination dependence of classical binaries. *Icarus* **194,** 758-768.

Peixinho, N., A. Doressoundiram, A. Delsanti, H. Boehnhardt, M.A. Barucci, and I. Belskaya 2003. Reopening the TNOs color controversy: Centaurs bimodality and TNOs unimodality. *Astron. & Astrophys.* **410,** L29-L32.

Peixinho, N., H. Boehnhardt, I. Belskaya, A. Doressoundiram, M.A. Barucci, A. Delsanti 2004. ESO large program on centaurs and TNOs: Visible colors - final results. *Icarus* **170,** 153-166.

Petit, J.M., J.J. Kavelaars, B. Gladman, and T. Loredo 2008. Size distribution of multikilometer transneptunian objects. In: A. Barucci, H. Boehnhardt, D. Cruikshank, A. Morbidelli (Eds.), *The Solar System Beyond Neptune*, Univ. of Arizona Press, Tucson, 71-87.

Sandford, S.A., and 54 co-authors 2006. Organics captured from Comet 81P/Wild 2 by the Stardust spacecraft. *Science* **314,** 1720-1724.

Simonelli, D.P., J.B. Pollack, C.P. McKay, R.T. Reynolds, and A.L. Summers 1989. The carbon budget in the outer Solar nebula. *Icarus* **82,** 1-35.

Stansberry, J., W. Grundy, M. Brown, D. Cruikshank, J. Spencer, D. Trilling, and J.L. Margot 2008. Physical properties of Kuiper belt and Centaur objects: Constraints from the Spitzer Space Telescope. In: A. Barucci, H. Boehnhardt, D. Cruikshank, A. Morbidelli (Eds.), *The Solar System Beyond Neptune*, Univ. of Arizona Press, Tucson, 161-179.

Tegler, S.C., W. Romanishin, and G.J. Consolmagno, S.J. 2003. Color patterns in the Kuiper belt: A possible primordial origin. *Astrophys. J.* **599,** L49-L52.

Tegler, S.C., J.M. Bauer, W. Romanishin, and N. Peixinho 2008. Colors of Centaurs. In: A. Barucci, H. Boehnhardt, D. Cruikshank, A. Morbidelli (Eds.), *The Solar System Beyond Neptune*, Univ. of Arizona Press, Tucson, 105-114.

Tholen, D.J. 1984. *Asteroid taxonomy from cluster analysis of photometry*. Ph.D. dissertation at Univ. of Arizona.

Thompson, W.R., B.G.J.P.T. Murray, B.N. Khare, and C. Sagan 1987. Coloration and darkening of methane clathrate and other ices by charged particle irradiation: Applications to the outer solar system. *J. Geophys. Res.* **92,** 14933-14947.

Vilas, F., and B.A. Smith 1985. Reflectance spectrophotometry (0.5-1.0 µm) of outer-belt asteroids: Implications for primitive, organic solar system material. *Icarus* **64,** 503-516.

Warren, S.G. 1984. Optical constants of ice from the ultraviolet to the microwave. *Appl. Opt.* **23,** 1206-1225.

Zellner, B., D.J. Tholen, and E.F. Tedesco 1985. The eight-color asteroid survey: Results for 589 minor planets. *Icarus* **61,** 355-416.